\documentclass[]{spie}  

\usepackage{amsmath,amsfonts,amssymb}
\usepackage{graphicx}
\usepackage[colorlinks=true, allcolors=blue]{hyperref}
\usepackage{subcaption}
\usepackage{needspace}

\title{Status and performance of the Gemini Planet Imager adaptive optics system}

\author[a]{Vanessa P. Bailey}
\author[b]{Lisa A. Poyneer}
\author[a]{Bruce A. Macintosh}
\author[c]{Dmitry Savransky}
\author[d]{Jason J. Wang}
\author[d]{Robert J. De Rosa}
\author[a]{Katherine B. Follette}
\author[b]{S. Mark Ammons}
\author[e]{Thomas Hayward}
\author[f]{Patrick Ingraham}
\author[g]{J\'{e}r\^{o}me Maire}
\author[b]{David W. Palmer}
\author[h]{Marshall D. Perrin}
\author[i]{Abhijith Rajan}
\author[e]{Fredrik T. Rantakyr\"o}
\author[f]{Sandrine Thomas}
\author[j]{Jean-Pierre V\'{e}ran}

\affil[a]{Kavli Institute for Particle Astrophysics \& Cosmology, Physics Department, Stanford University, Stanford, CA, 94305}
\affil[b]{Lawrence Livermore National Laboratory, 7000 East Ave, Livermore, CA, 94550, USA}
\affil[c]{Sibley School of Mechanical and Aerospace Engineering, Cornell University, Ithaca, NY 14853}
\affil[d]{Astronomy Department, University of California, Berkeley; Berkeley, CA 94720, USA}
\affil[e]{Gemini Observatory, Casilla 603, La Serena, Chile}
\affil[f]{Large Synoptic Survey Telescope, 950 N Cherry Ave, Tucson, AZ, 85719, USA}
\affil[g]{Dunlap Institute for Astronomy \& Astrophysics, University of Toronto, 50 St. George St., Toronto, Ontario, Canada}
\affil[h]{Space Telescope Science Institute, 3700 San Martin Drive, Baltimore, MD 21218, USA}
\affil[i]{School of Earth and Space Exploration, Arizona State University, PO Box 871404, Tempe, AZ 85287, USA}
\affil[j]{NRC Herzberg Institute of Astrophysics, 5071 W Saanich Rd, Victoria, BC V9E 2E7, BC, Canada}

\authorinfo{Further author information: (Send correspondence to V.P.B.)\\V.P.B.: E-mail: vpbailey@stanford.edu}

\begin{document}
\maketitle

\begin{abstract}
The Gemini Planet Imager is a high-contrast near-infrared instrument specifically designed to image exoplanets and circumstellar disks over a narrow field of view. We use science data and AO telemetry taken during the first 1.5~yr of the GPI Exoplanet Survey to quantify the performance of the AO system. In a typical 60~sec H-band exposure, GPI achieves a $5\sigma$ raw contrast of $10^{-4}$ at 0.4''; typical final $5\sigma$ contrasts for full 1~hr sequences are more than 10 times better than raw contrasts. We find that contrast is limited by bandwidth wavefront error over much of the PSF. Preliminary exploratory factor analysis can explain 60--70\% of the variance in raw contrasts with combinations of seeing and wavefront error metrics. We also examine the effect of higher loop gains on contrast by comparing wavefront error maps reconstructed from AO telemetry to concurrent IFS images. These results point to several ways that GPI performance could be improved in software or hardware.
\end{abstract}


\section{INTRODUCTION}
\label{sec:intro}  

Over the past two decades, our understanding of exoplanetary systems has progressed thanks to several complementary observational techniques. Transit and radial velocity surveys have yielded a wealth of new terrestrial planets, and a handful of the most tightly-orbiting ones can be characterized spectroscopically. However, no transit or radial velocity survey has a sufficiently long time baseline to detect, much less characterize, the planets in the outer reaches of extrasolar systems. Furthermore, our understanding of the formation and dynamical histories of these systems can be greatly aided by studying the morphology of the protoplanetary disks from which planets form and the debris disks that remain after the systems' evolution has stabilized. Spatially-resolved spectroscopy and polarimetry are powerful tools for investigating these systems.

High-contrast, high-resolution instrumentation is necessary to achieve these goals. To image even young, self-luminous gas giant exoplanets around their bright host stars requires near-infrared flux contrasts of $10^{-4} - 10^{-6}$. For even the nearest host stars, the projected separations of interest are often less than 0.5''.  The Gemini Planet Imager (GPI)\cite{Macintosh2008, Macintosh2014d} is a high-resolution, high-contrast near-infrared coronagraphic imager at Gemini South specifically designed to meet these requirements. GPI operates in one of two modes: a broadband polarimeter\cite{Perrin2010a} or an integral field spectrograph (IFS) with a wavelength-dependent spectral resolution of approximately 30--80\cite{Larkin2014}. Both modes have selectable bandpasses from Y to K-band, with pixel scales of $14.166 \pm 0.007$~mas/px and fields of view $2.7''$ square. High-contrast science is enabled by GPI's high-order adaptive optics system\cite{Poyneer2014}. For a thorough description and characterization of the AO system, as well as preliminary performance analysis based on 10~months of on-sky data, the reader may refer to Poyneer et al. (2016)\cite{Poyneer2016}.  In these proceedings, we provide an updated description the AO system performance, particularly as related to science images, as well as lessons learned, based on data from the first 1.5~yr of the GPI Exoplanet Survey (GPIES).

\section{ADAPTIVE OPTICS SYSTEM DESCRIPTION}
\label{sect:GPIdescription}

GPI AO is a natural guide star system with a woofer/tweeter mirror pair and a spatially filtered Shack-Hartmann wavefront sensor (SH WFS) \cite{Poyneer2006a, Poyneer2014, Poyneer2016}. The woofer is a nine actuator diameter piezo mirror from Cilas, with a 5~mm actuator pitch. The tweeter is a  micro-electromechanical system (MEMS) device from Boston Micromachines; it has an illuminated diameter of 43.2 actuators with a 0.4~mm actuator pitch. The tweeter actuators are one-to-one mapped to the quad-cell SH WFS subapertures. The 160$\times$160~px CCID-66 sensor operates at $\sim$700--900~nm, with selectable bandpass filters for the brightest stars. The WFS has an adjustable-width spatial filter that is used to reduce the effects of aliasing\cite{Poyneer2004}. The control loop is run at 1~kHz on bright stars ($I<8$), and at 500~Hz on fainter stars. Due to WFS detector read noise limitations, we do not observe stars fainter than $I\sim9.5$. GPI AO uses an LQG controller\cite{LeRoux2004,Poyneer2010} to mitigate tip, tilt, and focus vibrations at 60~Hz (and harmonics). An additional vibration notch can be added if necessary, to counteract other sources of telescope vibration. LQG controllers are also employed by SPHERE\cite{Petit2014} and CANARY\cite{Sivo2014}.

\begin{figure}[tbh]
\begin{center}
\includegraphics[width=6in]{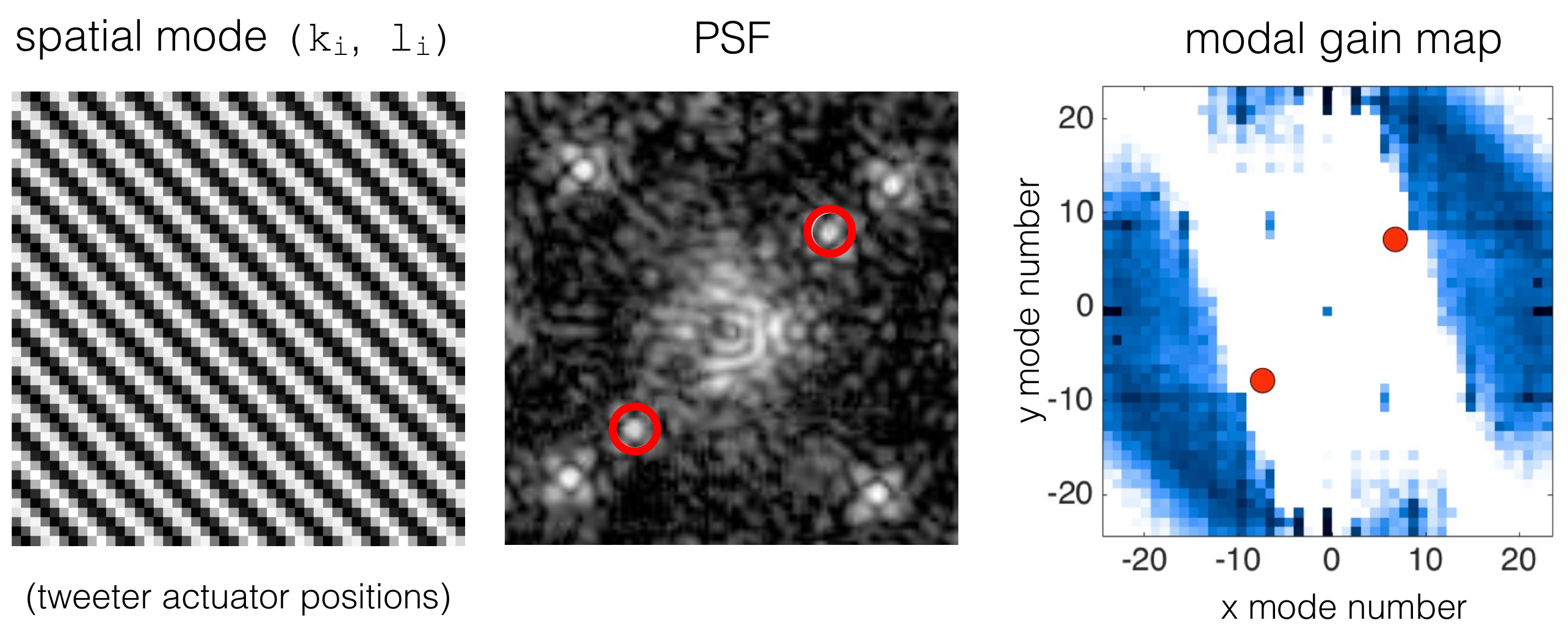}
\end{center}
\caption{ \label{fig:FTR}
Correspondence between spatial mode, PSF location, and gain optimizer map. \textit{Left:} Tweeter commands for an example Fourier mode. \textit{Middle:} This Fourier mode maps to a single pair of spots in the PSF. Image field of view is matched to the DM control region. \textit{Right:} 2D gain map, with the same Fourier mode marked. Piston (mode 0,0) is at the center.}
\end{figure}

GPI uses a Fourier Transform Reconstructor (FTR) with independently optimized gains\cite{Poyneer2005}. In addition to computational efficiency, the FTR has the added benefit of simple mapping between spatial modes and location in the PSF. A sinusoidal phase error maps to a pair of points (and their harmonics) in the focal plane; in the high Strehl Ratio (SR) case, only the primary pair of points dominate. The gain is optimized on a mode-by-mode basis, as described below. Figure \ref{fig:FTR} shows the mapping between spatial mode, PSF location, and gain map location.

\begin{figure}[tbh]
\begin{center}
\includegraphics[width=6.5in]{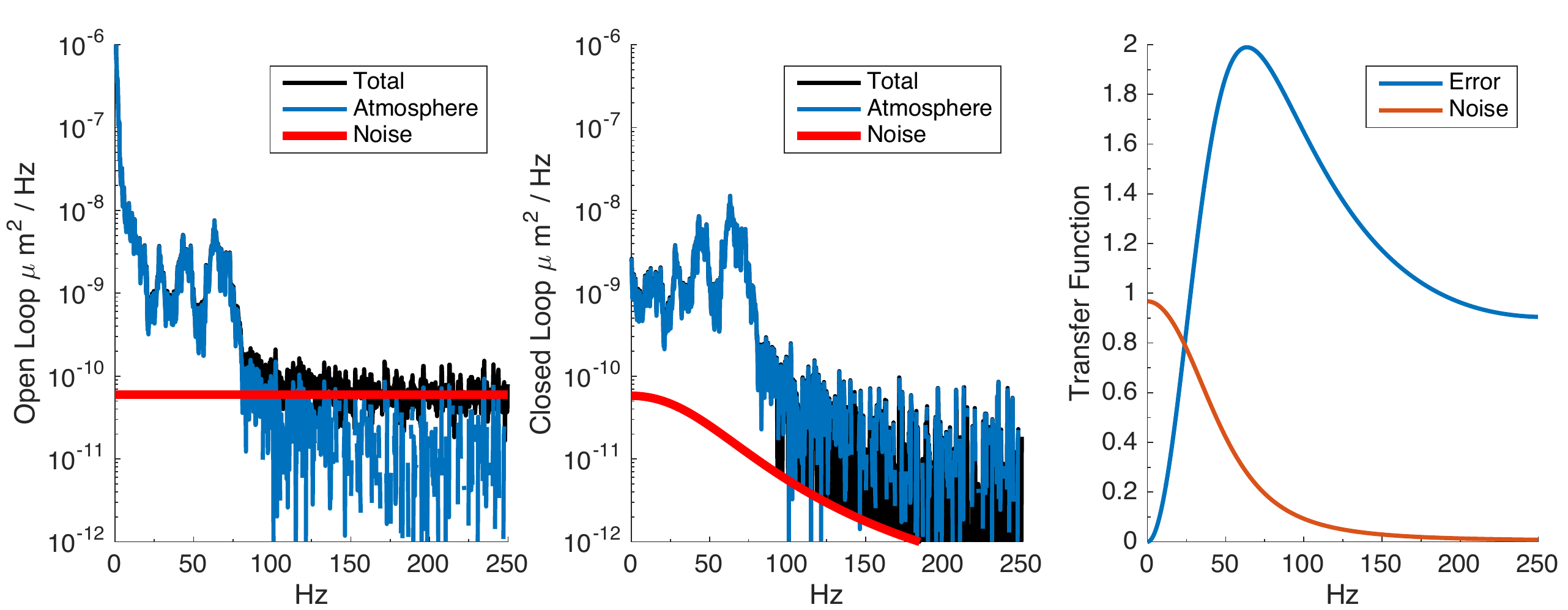}
\end{center}
\caption{ \label{fig:PSD_TF}
\textit{Left:} Reconstructed open-loop PSD for a single Fourier mode. The black line is the total PSD, the red line is the noise PSD, and the blue line (total--noise) is the atmosphere PSD.
\textit{Center:} Corresponding reconstructed closed loop PSDs at the science leg.
\textit{Right:} ETF and NTF for the gain chosen for this mode.
}
\end{figure}

GPI uses Power Spectral Density (PSD) analysis for modal gain optimization. The PSD at the WFS is constructed for each mode  from a timeseries of measured phases using a modified periodgram analysis. Then, by dividing by the theoretical Error Transfer Functions (ETFs) for each mode and gain, we  reconstruct theoretical open loop PSDs for each mode (Figure \ref{fig:PSD_TF}, left panel). The theoretical closed loop PSD in the science arm (Figure \ref{fig:PSD_TF}, center panel) is calculated from the open loop PSD by multiplying by the appropriate ETF and Noise Transfer Function (NTF) for the mode and gain applied\footnote{The ETFs and NTFs for each mode have been verified off-sky with an artificial source.} (Figure \ref{fig:PSD_TF}, right panel). Unless otherwise stated, we refer to closed loop PSDs at the science arm, not the WFS arm.  For the remainder of the paper, we use the terms ``bandwidth wavefront error'' ($\sigma_{BW}$) and ``noise wavefront error'' ($\sigma_{noise}$) to refer to the sums of the atmosphere PSD and noise PSD, respectively, delivered to the science arm. Note that by definition, we cannot calculate PSDs for spatial frequencies above our Nyquist-sampling limit, so their contribution (i.e.: fitting error and aliasing) is not captured in the total bandwidth error.

The results of these PSD analyses are used to optimize gains on a mode-by-mode basis every 8~sec. Higher gains more strongly attenuate low temporal frequency signals, at the expense of greater overshoot at higher temporal frequencies and greater noise contribution. Using a root-finding algorithm, the optimizer chooses the gain that minimizes the closed loop power (bandwidth + noise power), effectively minimizing the wavefront error at the science arm. A gain cap of 0.3 is currently imposed based on stability analysis. The optimization process is further described in previous work\cite{Poyneer2005, Poyneer2016}.

In practice, GPI operation is non-ideal in several ways. First, four tweeter actuators within the illuminated pupil are dead and are blocked as part of a custom apodized-pupil Lyot coronagraph mask\cite{Soummer2005}. Second, for unknown reasons, we find that we cannot close the spatial filter fully to the radius corresponding to the Nyquist-sampled spatial frequency of the wavefront sensor (the ``control radius''), due to flux loss and loop instabilities. Therefore, we typically operate the spatial filter 50\% oversized in median conditions. Third, the non-common path aberration calibration subsystem\cite{Macintosh2008} only functions in the lab, due to yet-unresolved vibration sources encountered when mounted on the telescope\cite{Macintosh2016a}. Nonetheless, NCPA do not dominate our error budget\cite{Poyneer2016}. Finally, because GPI's WFS is a quad-cell SH design, it is prone to so-called ``centroid gain'' variability. If the PSF quality is degraded, and the spatial filter cannot be closed tightly, the spot size on the quad cell grows beyond $\lambda/d_{subap}$. Hence, for the same spot displacement, a smaller slope is measured, decreasing the effective gain of the system. We cannot yet reliably measure the system centroid gain; in the meantime we can attempt to compensate for it, as described in Section \ref{sec:gain_opt_cap}.

\section{DATASETS AND REALTIME PROCESSING}
One of the major strengths of GPI and the GPIES campaign is realtime science data processing and AO performance monitoring. Images are processed in real time, and headers include a variety of diagnostic information and performance metrics. Additionally, we can manually save detailed AO telemetry sets. Finally, header and AO telemetry metadata from GPIES campaign observations are aggregated in a mySQL database \cite{Perrin2016}. In the first 1.5~yr of the GPIES campaign, a total of 318 H-band IFS observing sequences were taken, for a total of more than 10,000 science frames. Of these,  $\sim1100$ frames (249 sequences) had a set of detailed AO telemetry taken during the frame. The overwhelming majority of these images were taken in standard campaign mode (60~sec exposures), typically with 20--40 images per sequence. Unless otherwise noted, only the subset of  IFS data with matched AO telemetry is presented.

Every science frame is processed in real time. In IFS mode the microspectra are extracted to produce a 3D $x/y/\lambda$ cube, while in polarimetry mode the two orthogonal polarization directions are extracted to produce an $x/y/pol$ cube. These cubes are then high-pass filtered and ``quicklook'' processed to produce single-frame $5\sigma$ contrast curves at each wavelength/polarization, and the average $5\sigma$ contrast across all wavelengths or polarizations at 0.25'', 0.4'', and 0.8'' is saved in the image header\cite{Perrin2016}. We refer to these single-frame contrasts as ``raw contrasts.'' Note that throughout these proceedings we quote $5\sigma$ contrasts, because that is the typical signal-to-noise threshold used for exoplanet detection; $1\sigma$ contrast curves (the ``noise floor'') are correspondingly 5 times lower. An approximate measure of the average wavefront error (WFE) is also saved with each frame. However, this value is estimated only from the temporal variation of the centroids, without wavefront reconstruction or correction for the ETF, and is systematically 20--70~nm higher than the WFE reconstructed from full AO telemetry sets. Hence, all wavefront error values quoted in this proceedings are derived from full AO telemetry analysis.

Manually triggered AO data sets provide a detailed view of AO performance. Each telemetry set includes: all of the tweeter commands, woofer commands, reconstructed phase residuals, and tip/tilt data; a time-decimated set of 2D gain maps, WFS images, and reference centroids; and the control parameters. These data are reduced with custom IDL scripts to produce reconstructions of quantities such as the tip/tilt vibration amplitude, PSDs of closed and open loop data, and contributions from noise and bandwidth WFEs\cite{Poyneer2014}. Typically, 22~sec telemetry sets are taken 1--5 times per target; data are not taken with every science frame, because the full data rate is $>1.1$~GB/min.

Every science and AO header also contains various environmental parameters such as windspeeds (ambient and at the secondary mirror) as well as seeing estimates and turbulence profiles from the observatory's Differential Image Motion Monitor (DIMM) and Multi-Aperture Scintillation Sensor (MASS)\cite{Sarazin1990, Kornilov2003}. Approximately half of our matching IFS+AO telemetry datasets ($\sim500$) had valid DIMM and MASS measurements. Stale DIMM/MASS measurements were culled if the values were constant for at least 30 minutes, because that is currently our best available indication for when the MASS/DIMM is closed or otherwise nonoperable. Future work will focus on reconstructing $r_0$ and $\tau$ from AO telemetry itself; this will be a more reliable indicator of the turbulence actually seen by GPI.

\section{ANALYSIS AND RESULTS}

\subsection{Contrast performance}
\begin{figure}[ht]
	\begin{center}
   \includegraphics[width=5in]{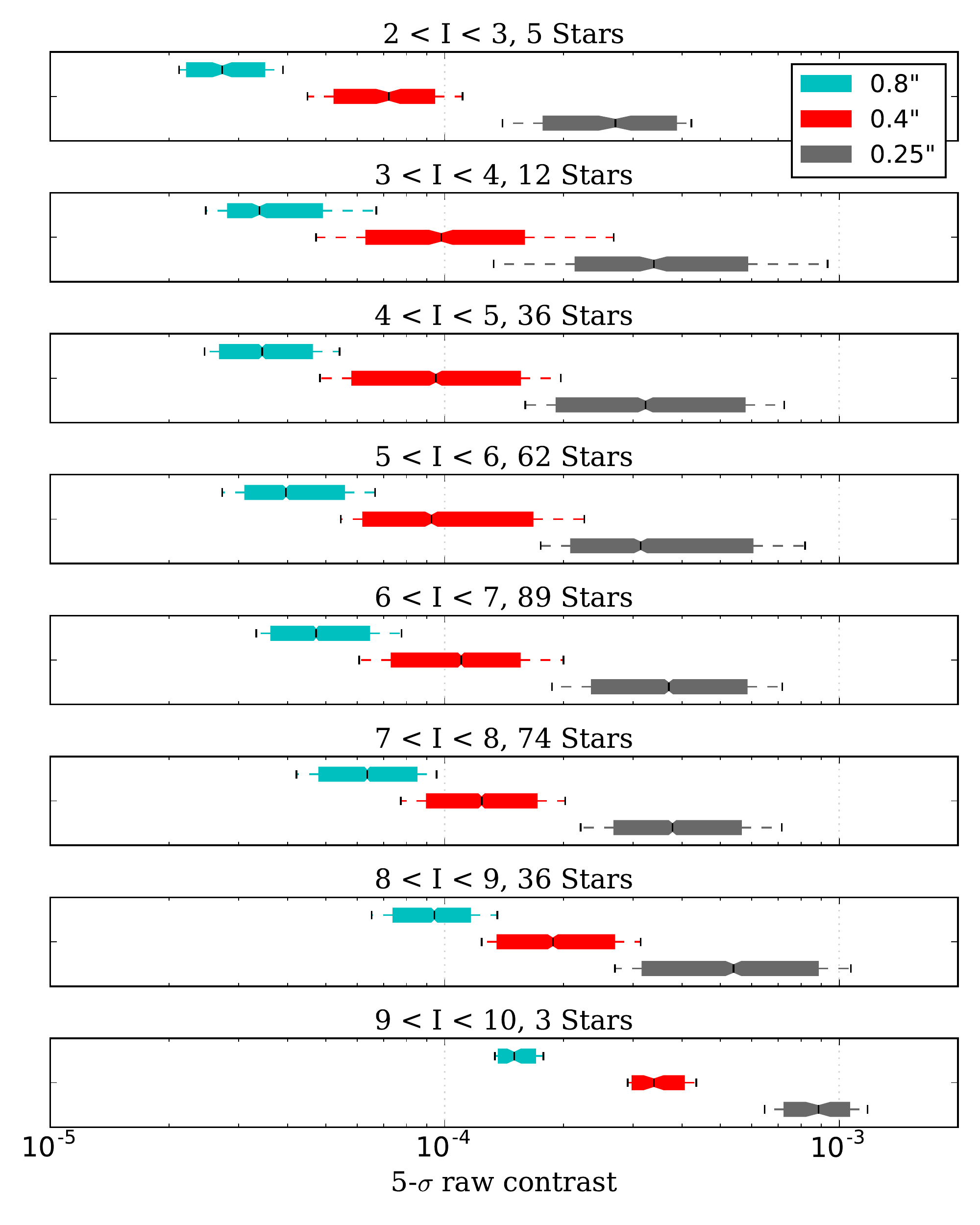}
	\end{center}
   \caption{ \label{fig:raw_boxplot} Distribution of $5\sigma$ raw contrasts for all 1.65~$\mu$m coronagraphic IFS images. For each range of stellar I-band magnitude, the contrasts at three projected separations are shown: blue = 0.25'', red = 0.4'', and gray = 0.8''.  The median of each distribution is marked by the notch and vertical line in the box; boxes extend from the the 25th percentile to the 75th percentile; whiskers extend from the 16th percentile to the 84th percentile. Note that the AO framerate switches from 1~kHz to 500~Hz at $I=8$.}
\end{figure}

We first investigate GPI's raw contrast performance. The distributions of $5\sigma$ raw contrasts for all $1.65~\mu$m coronagraphic IFS frames in the GPIES campaign, subdivided by projected separation and guide star I-band magnitude, are shown in Figure \ref{fig:raw_boxplot}. Several trends are apparent. First, contrasts at all projected separations worsen for $I>8$ guide stars. This is because, in addition to IFS background noise (read noise plus sky and thermal backgrounds), the AO system switches from 1~kHz to 500~Hz at $I=8$. Second, contrast at 0.25'' is roughly independent of stellar magnitude for $I<8$, because it is determined by AO residuals, not by the background limit. The effects of background noise become apparent at 0.4'' only at $I\sim7$. At 0.8'', under good conditions, images may be background-limited on even brighter targets.

\begin{figure}[ht]
    \begin{center}
    \includegraphics[width=\textwidth]{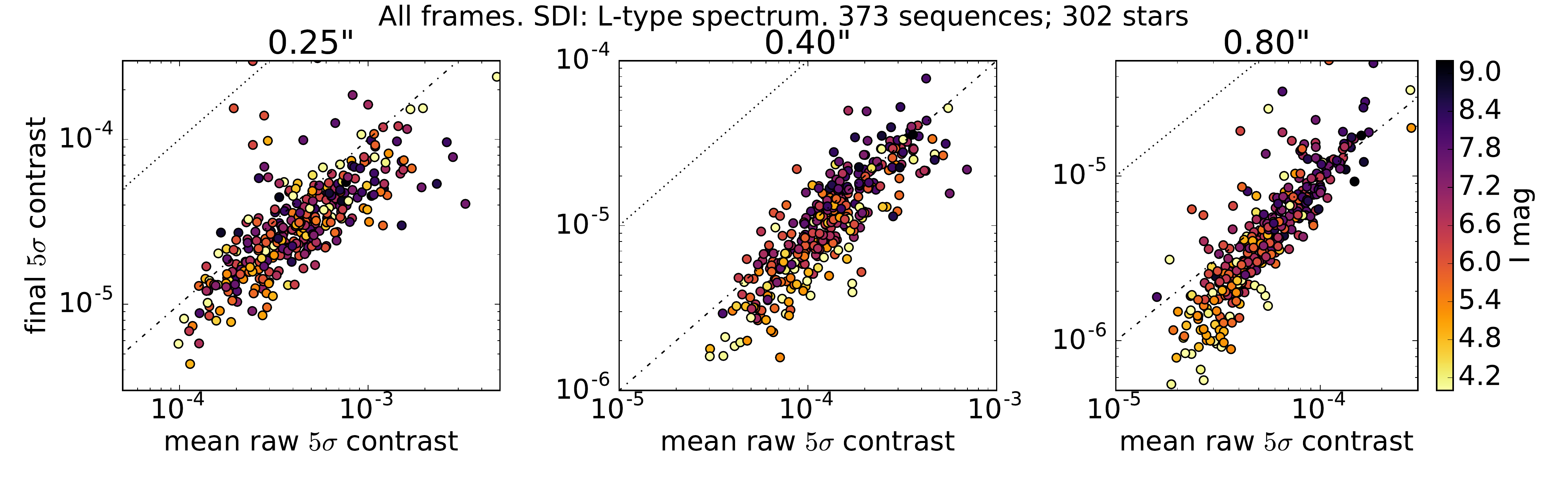}
    \end{center}
   \caption{ \label{fig:raw_vs_final} Final H-band $5\sigma$ contrast (assuming a conservative, ``flat'' L-type planet spectrum for the SDI reduction) vs.\ mean raw $5\sigma$ contrast. Points are color-coded by the guide star magnitude. Dotted line is a 1:1 ratio, dot-dashed line is a 1:10 ratio.}
\end{figure}

Final $5\sigma$ contrasts for the $1.65~\mu$m coronagraphic sequences are shown in Figure \ref{fig:raw_vs_final}. These data were processed using both angular and spectral differential imaging (ADI/SDI), with KLIP PSF modeling and subtraction\cite{Marois2006, Soummer2012, Wang2015b}. A ``flat'' L-type planet spectrum was assumed for the final contrasts shown; an additional factor of $\sim2$ in contrast is typically gained if a strongly-peaked T-type planet spectrum is assumed instead. Final ``flat'' contrasts are typically at least 10 times more sensitive than raw contrasts. As with raw contrast, the final contrast at 0.8'' depends strongly on guide star magnitude, because it is limited by background noise, while final contrast at 0.25'' is nearly independent of guide star magnitude.

\begin{figure}[ht]
	\begin{center}
   \includegraphics[width=6.5in]{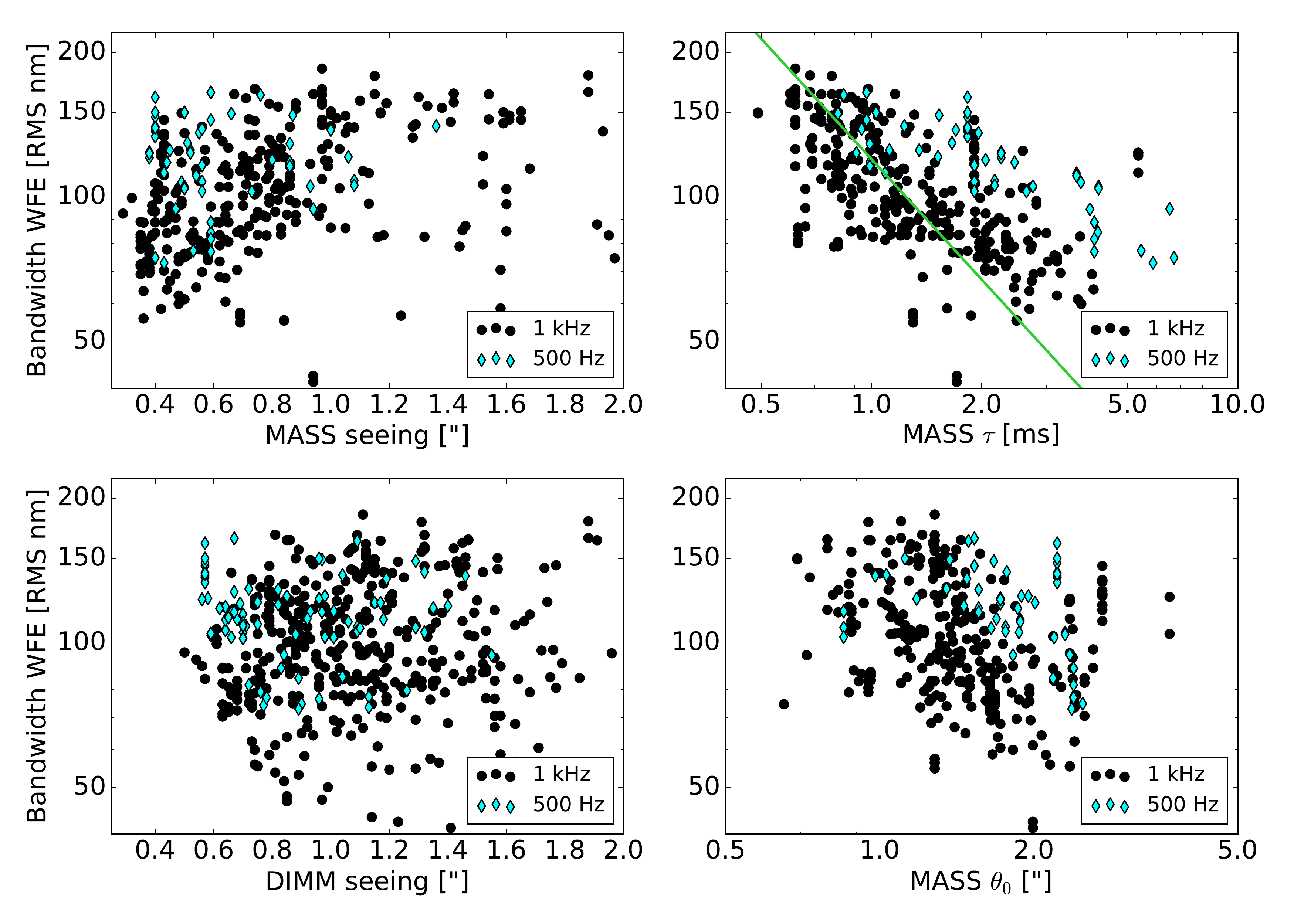}
	\end{center}
   \caption{ \label{fig:wfe_seeing} Reconstructed AO bandwidth error vs. several seeing parameters reported by the DIMM and MASS. Black points are 1~kHz AO frame rate, while the blue diamonds are 500~Hz. Green line is a $\sigma_{BW}^2 \propto \tau^{5/3}$ trend line.}
\end{figure}

\clearpage
\subsection{AO performance dependence on atmospheric conditions}

We next study correlations between bandwidth WFE and atmosphere parameters delivered by the MASS/DIMM (Figure \ref{fig:wfe_seeing}). Our last published analysis of these correlations \cite{Poyneer2016} included only the first 10 months of campaign data, and did not filter out stale seeing measurements. However, the observed relationships remain similar.  Bandwidth WFE is not strongly correlated with the seeing measurements from the DIMM. However, at low significance, there may be a lower limit to the wavefront error achievable for a given value of MASS seeing. There is a weak trend with isoplanatic angle, $\theta_0$; however, $\theta_0$ is degenerate with the height of the turbulent layers. The relationship to the characteristic timescale of seeing ($\tau = 0.31 r_0 / v_{wind}$) does follow the expected $\sigma^2 \propto \tau^{-5/3}$ power law, albeit with significant scatter. In particular, WFE at the shortest $\tau$ is better than expected from a fit to longer $\tau$. This could indicate either a systematic error in the measurement of very short $\tau$ or an under-performing system at the longest $\tau$. As we will demonstrate, bandwidth error due to wind lag is the primary factor limiting GPI's performance on bright stars. GPI was designed for an assumed median $\tau$ of $\sim5$~ms\cite{Poyneer2016, Poyneer2006}, while the average MASS $\tau$ is $<2$~ms.


\subsection{Combined AO and IFS analysis}
\label{subsect:AO+IFS}

We next seek a deeper understanding of the link between AO parameters and science data quality with a joint analysis of the two datasets. Preliminary versions of some of these analyses, based on early GPIES campaign data, were previously published\cite{Poyneer2016}. Our analysis is based on 1071 raw IFS frames with AO telemetry; we have a total of 90 observing sequences with MASS data, 113 observing sequences with DIMM data, and 246 observing sequences with AO telemetry.

\begin{figure}[ht]
	\begin{center}
   \includegraphics[width=6in]{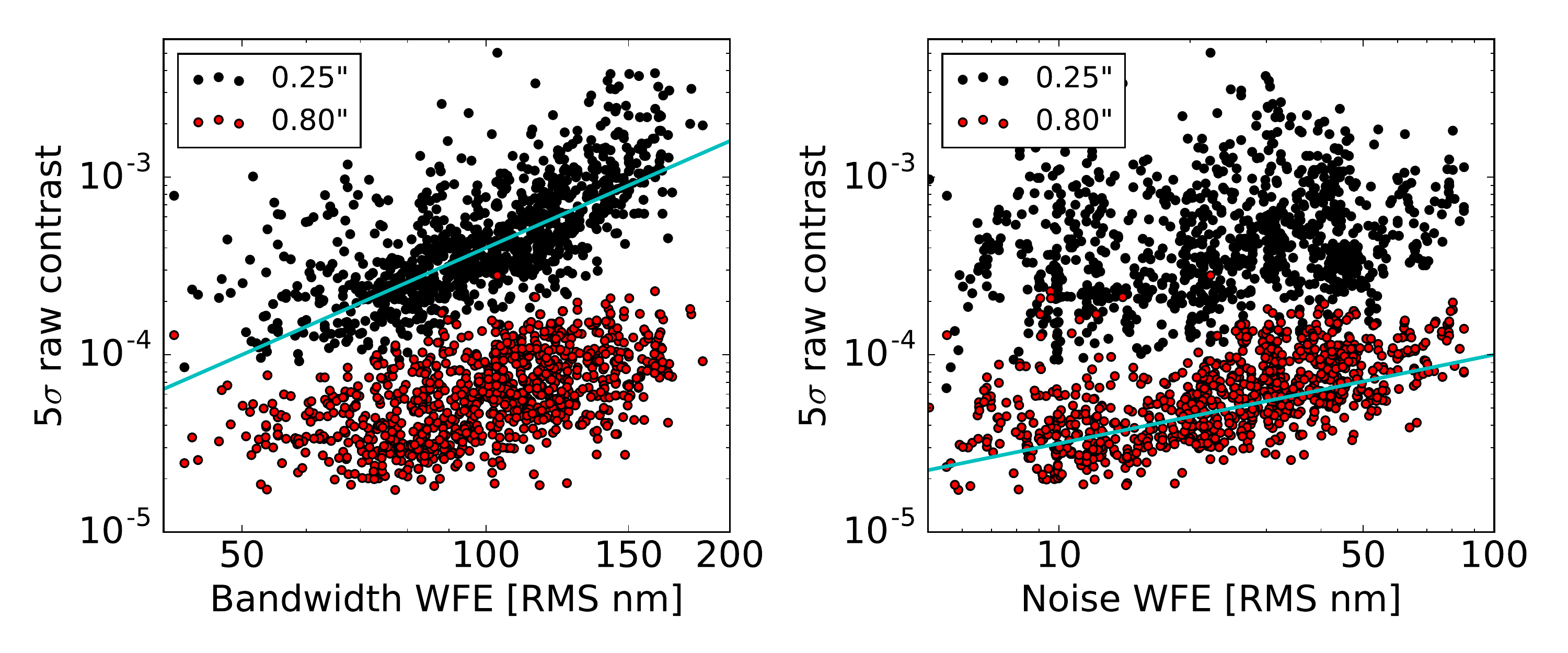}
	\end{center}
   \caption{ \label{fig:raw_contrast_vs_WFE} Raw $5\sigma$ contrast at $1.65~\mu$m   vs.\ bandwidth and noise WFE terms. Contrasts for two separations are shown: black = 0.25'' and red = 0.8''. The contrast at 0.25'' is dominated by atmosphere residuals ($\propto \sigma_{BW}^2$), while contrast at 0.8'' is background-limited ($\propto \sigma_{noise}^{1/2}$). Blue lines indicate these two expected trendlines.}
\end{figure}

Raw contrast is a clear function of WFE (Figure \ref{fig:raw_contrast_vs_WFE}). At small separations, bandwidth error residuals (speckles) dominate; the 0.25'' raw contrast follows a $\sigma_{BW}^2$ trend, while it is nearly uncorrelated with $\sigma_{noise}$. At larger separations, background noise dominates, with a smaller contribution from bandwidth error.  The 0.8'' raw contrast follows a $\sigma_{noise}^{1/2}$ trend, although there is scatter in this relationship due to a sub-dominant contribution from $\sigma_{BW}$.

\begin{figure}[ht]
	\begin{center}
   \includegraphics[width=5.5in]{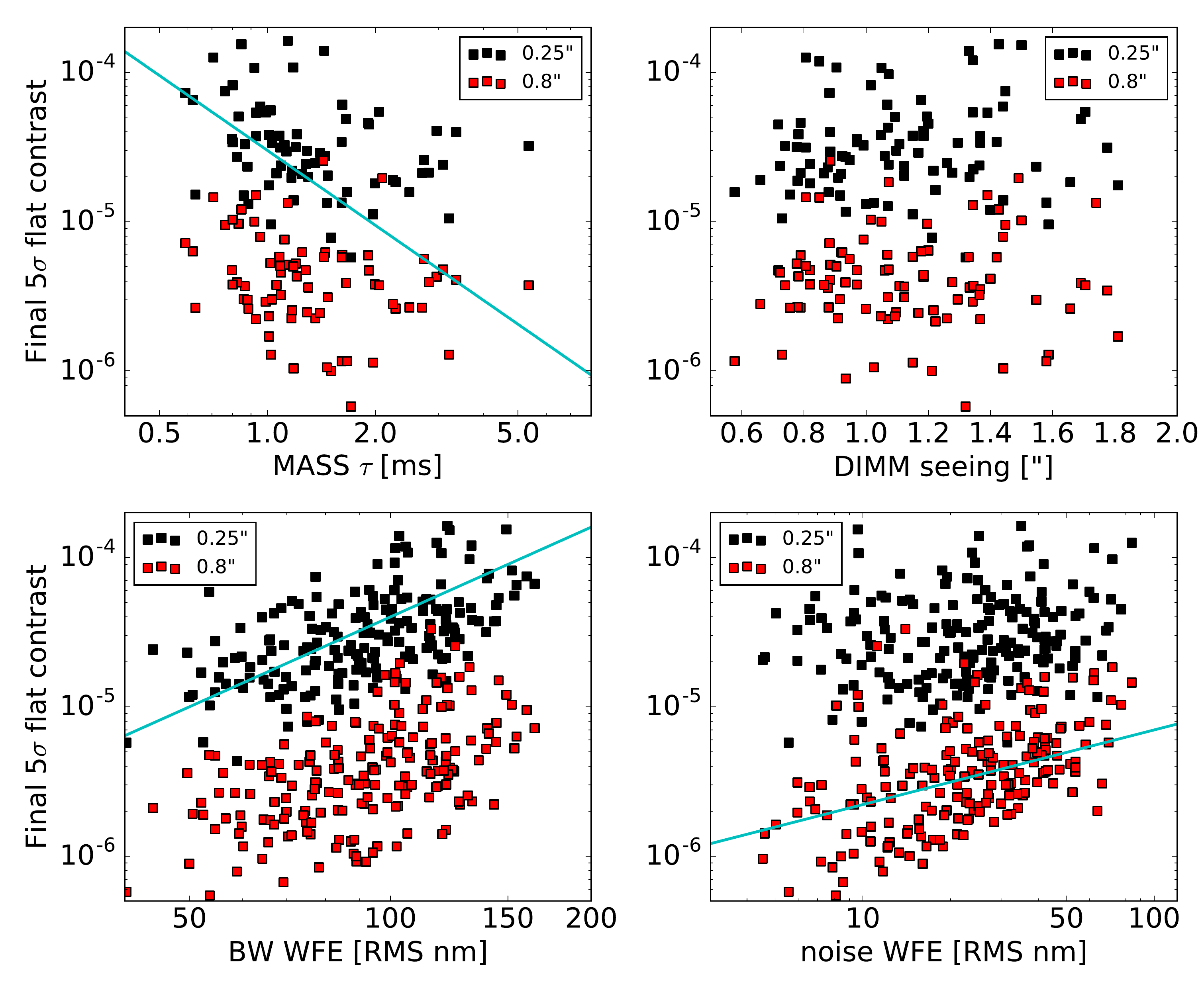}
	\end{center}
   \caption{ \label{fig:final_contr_1000_vs_WFE} Final $5\sigma$ contrast at $1.65~\mu$m  vs. seeing parameters and reconstructed WFE terms. Contrasts for two separations are shown: black = 0.25'' and red = 0.8''. Final contrast is  uncorrelated with DIMM seeing. Contrast in the inner regions is still dominated by bandwidth error, while the outer regions are now governed by a combination of bandwidth error and background noise. Blue lines are $\tau^{5/3}$, $\sigma_{BW}^2$, and $\sigma_{noise}^{1/2}$ trendlines.}
\end{figure}

Final contrast similarly depends on WFE and MASS $\tau$, while it is uncorrelated with DIMM seeing (Figure \ref{fig:final_contr_1000_vs_WFE}). Final contrast at 0.25'' is still dominated by $\sigma_{BW}$. Final contrast at 0.8'' has a much stronger contribution from $\sigma_{BW}$ than did raw contrast, because integration time has beaten down the background noise. As a result, 0.8'' final contrast does not follow a $\sigma_{noise}^{1/2}$ trend.  With fewer datapoints, the relationship to MASS $\tau$ is less evident than it was for raw contrast. However, the same degradation in contrast with decreasing $\tau$ is still apparent, as we expect given the dependence of contrast on $\sigma_{BW}$.

\begin{figure}[ht]
	\begin{center}
   \includegraphics[width=4.75in]{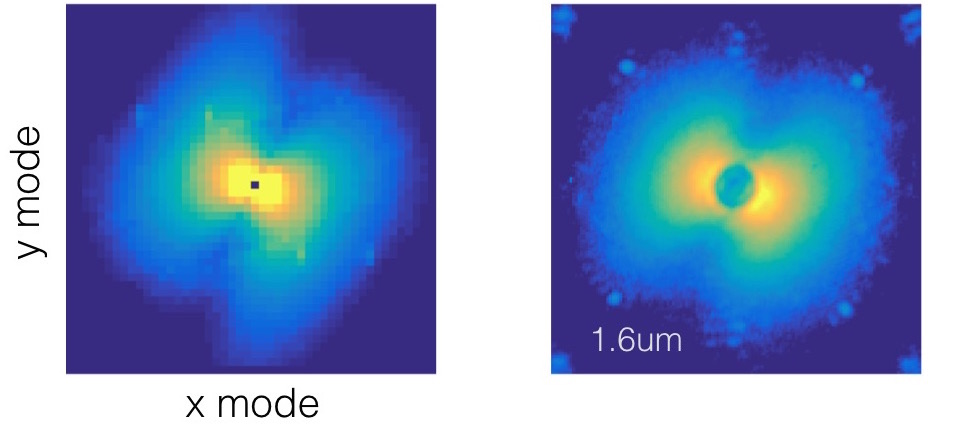}
	\end{center}
   \caption{ \label{fig:WFE_PSF_comp} Comparison of bandwidth WFE per mode and a simultaneous IFS exposure over the corresponding field of view. In the bandwidth error-dominated regime, the PSF intensity scales as the square of the RMS bandwidth WFE in the corresponding Fourier mode.  In the IFS image, the central dark circle is due to the coronagraph, and the four spots just outside the wind butterfly are injected by a pupil plane grid in the science camera.}
\end{figure}

We can also verify the relationship between focal plane images and reconstructed WFE in two dimensions. From the analysis above, we see that contrast across most of the PSF is proportional to the square of the bandwidth WFE, so we compare reconstructed bandwidth WFE maps to IFS frames. In these maps, the total WFE in each mode is the integrated power under each  bandwidth error PSD. Because GPI uses a Fourier basis set for modal control, the RMS WFE in a given mode can be directly mapped to the corresponding location in the PSF. Figure \ref{fig:WFE_PSF_comp} shows one such example. Qualitatively, the morphology of the PSF is well-matched to the reconstructed WFE\footnote{The WFE map is by definition missing contributions from non-common path errors, downstream optics in the science camera, and WFS aliasing.}. Such comparisons not only verify our understanding of the process of WFE reconstruction, but can guide future targeted efforts for contrast improvement, and may potentially inform PSF subtraction. More thorough studies of PSF reconstruction from AO telemetry are left to future work.

\Needspace{10\baselineskip}
\subsection{Exploratory factor analysis of raw contrast}
\label{subsect:EFA}

A long term goal for the GPI instrument is to be able to predict raw and final contrast, given information about current conditions and/or AO performance. Gemini typically operates in queue mode, and GPIES observations are executed in priority visitor mode; in both cases, observers need to decide whether conditions are of sufficient quality to meet the science goals of a given sequence. If we can predict final contrast for a sequence, with reasonable error bars, we can help observers choose when to execute GPI sequences. To this end, we present some preliminary results from exploratory factor analysis of raw contrast data.

Exploratory Factor Analysis (EFA) is a tool designed to identify underlying latent variables based on common behavior in a number of measured variables. Given a set of independent variables, EFA groups them into ``factors'' based on the degree of correlation between the variables' behavior. Computed latent variables (factors) are believed to be a more robust measure of a complex underlying phenomenon (e.g. atmospheric conditions) than can be captured with a single measured variable alone (e.g. DIMM seeing measure). In our case, the measured variables considered in this initial analysis include all AO telemetry metadata and atmospheric condition measurements. As the purpose of this EFA analysis was to identify more robust predictors of contrast, and since we expect these variables to have power law relationships with contrast, all measured variables were log-scaled before EFA.

We completed our EFA analysis using IBM's SPSS Statistics software's\cite{IBM2015a} built-in ``dimension reduction'' capabilities\footnote{For an accessible introduction to Factor Analysis and choices of methods and parameters, including all that are reported/chosen here, see Discovering Statistics Using IBM SPSS, Chapter 17 (Field, 2013)}. We chose principal factors analysis\footnote{similar in principle to Principal Component Analysis, which astronomers are likely more familiar with, but without the assumption of common variance.}, done using the ``maximum likelihood'' method in SPSS and the oblique rotation method ``direct oblimin,'' which allows the extracted factors to be correlated with one another rather than forcing them to be orthogonally related.

Based on the initial output, we removed several variables from analysis that were either redundant (colinear) or poorly sampled. We define redundant variables based on off-diagonal correlation matrix values $>$0.9, and remove all but one of the redundant variables in each case. An example redundant pair is H-band and I-band magnitudes; we kept I-band magnitude.  We also removed total WFE, because both bandwidth WFE and noise WFE were included.  We also removed two variables, MASS $C_N^2$ at 0.5~km and the centroid gain estimate, whose diagonal values in the Anti-Image Correlation Matrix indicated a lack of sampling adequacy ($<0.5$).

The number of factors to extract was determined based on a scree plot, which shows a marked inflection at five factors, suggesting that a four factor solution was in order (indeed, factor solutions with $n_{factor}\geq5$ do not converge for this dataset).  Table \ref{tab:EFA_Table} shows the final set of 17 independent variables used in our EFA, and the resulting factor loading scores for the four extracted factors. These four factors appear to describe: (1) upper atmosphere conditions, (2) noise errors, (3) bandwidth errors, and (4) lower atmosphere conditions. Two variables, MASS $C_N^2$ at 8~km and 4~km, did not load onto any factors.

\begin{table}[htb]
\begin{center}
\begin{tabular}{|l | c | c | c | c | l |}
\hline
Variable (log-scaled) & F1 &  F2  &  F3 & F4 & GPI variable name\\
\hline
MASS $\theta_0$  &-1.016 & & & & MASSISOP \\
$C_N^2$ 16~km    & 0.869 & & & & MASS16CN \\
MASS seeing      & 0.679 & & & 0.545 & MASSSEE\\
MASS $\tau$      &-0.662 & & & & MASSTAU\\
WFS flux         & &-0.996 & & & CTPERSUB \\
noise WFE        & & 0.981 & & & WFE\_NOISE\\
star I mag     & & 0.827 & & & IMAG \\
AO frame rate    & &-0.494 & & & AOFRAMES\\
X tip/tilt error     & & & 0.954 & & EX \\
Y tip/tilt error     & & & 0.912 & & EY \\
bandwidth WFE    & & & 0.617 & & WFE\_BAND \\
focus vibration  & & & 0.599 & & WFE\_FOC \\
$C_N^2$ 2~km     & & & & 0.695 & MASS2CN2 \\
$C_N^2$ 1~km     & & & & 0.655 & MASS1CN2 \\
DIMM seeing      & & & & 0.475 & DIMMSEE \\
\hline
$C_N^2$ 4~km     & & & & & MASS4CN2\\
$C_N^2$ 8~km     & & & & & MASS8CN2\\
\hline
\end{tabular}
\caption{ \label{tab:EFA_Table}  Independent variables and resulting EFA factor loadings. Only factor loadings above 0.3 are shown. Two variables did not load onto any factors. All independent variables were log-scaled before analysis. The factors broadly describe (1) upper atmosphere conditions, (2) noise errors, (3) bandwidth errors, and (4) lower atmosphere conditions. }
\end{center}
\end{table}

\begin{figure}[tbh]
	\begin{center}
   \includegraphics[width=6.75in]{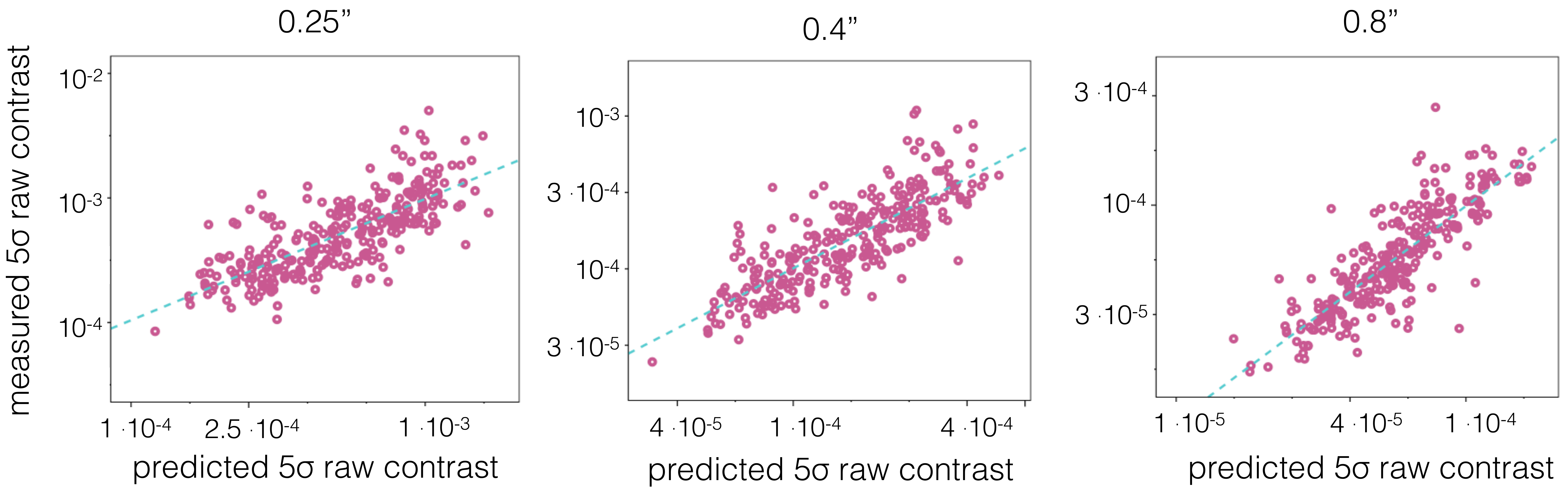}
	\end{center}
   \caption{ \label{fig:EFA}  Comparison of measured and predicted $5\sigma$ raw contrasts at 0.25'', 0.4'', and 0.8''. Linear regression parameters are listed in Table \ref{tab:EFA_fits}. Blue dashed line is 1:1 correspondence.}
\end{figure}

We then engaged in a linear regression analysis to predict raw contrast values based on all four factors as well as the variables that we had previously removed from the EFA analysis due to redundancy or poor sampling. Table  \ref{tab:EFA_fits} shows the linear combination of factors that describes the maximum amount of variance in the raw contrast at each separation, and Figure \ref{fig:EFA} shows the resulting fits. In all cases, standardized regression (beta) coefficients are listed only for variables that show a significant ($p<0.05$) linear relationship with log contrast at that separation. In all three cases, a linear regression model generated from the four EFA factors alone has an $r^2$ value of $\sim0.3$. The inclusion of additional standalone variables in the regression model, particularly the tip/tilt bandwidth error and total WFE, doubles the amount of explained raw contrast variance to between 60--70\%.

\begin{table}[htb]
\begin{center}
\begin{tabular}{|l | c | c | c | l |}
\hline
Variable & 0.25'' & 0.4'' & 0.8'' & GPI variable name\\
\hline
F1 (upper atm.) & & & 0.218 & \\
F2 (noise)      & & & 0.296 & \\
F3 (bandwidth)  & 1.333 & 1.549 & 2.508 & \\
F4 (lower atm.) & & 0.098 & & \\
star H mag   & & & 0.256 & HMAG\\
X tip/tilt BW$^+$ & -0.806 & -1.052 & -1.743 & EXBAND \\
Y tip/tilt BW & -0.448 & -0.421 & -0.607 & EYBAND \\
total WFE: $\sigma_{BW} + \sigma_{noise}$ & 0.461 & 0.499 & & WFE\\
centroid gain$^*$ & 0.343 & 0.287 & 0.245 & CGAINADJ \\
\hline
R$^2$ & 0.592 & 0.667 & 0.727 & \\
\hline
\end{tabular}
\caption{ \label{tab:EFA_fits} Standardized beta coefficients for linear regression fits to raw contrast. Additional independent variables were log-scaled. A blank entry indicates that the relationship was not significant (p$>$0.05). \\
$^+$ The bandwidth contribution to tip/tilt error is analogous to the bandwidth contribution to WFE.\\
$^*$ GPI's centroid gain estimate is approximate, and hence has weak predictive power.}
\end{center}
\end{table}

The regression model fits show many of the trends expected given the correlations discussed in Section \ref{subsect:AO+IFS}, with some differences in need of further study. As expected, contrast at 0.25" is fit primarily by variables related to bandwidth error. Also as expected, contrast at 0.8" requires terms related to noise limits in the AO system and the science camera (F2 and star H magnitude), and does not require the standalone total WFE term. The physical interpretation of the effects of the lower and upper atmosphere components at 0.8'' and 0.4'' will require further study. Additional work is also needed to understand why redundant variables  (tip/tilt bandwidth error and total WFE) are not adequately captured in the extracted factors and must be added as standalone variables. Future work will refine this preliminary analysis of raw contrasts, and expand the analysis to final contrasts as well.


\Needspace{8\baselineskip}
\subsection{Gain optimizer cap}
\label{sec:gain_opt_cap}

\begin{figure}[tbh]
	\begin{center}
   \includegraphics[width=5in]{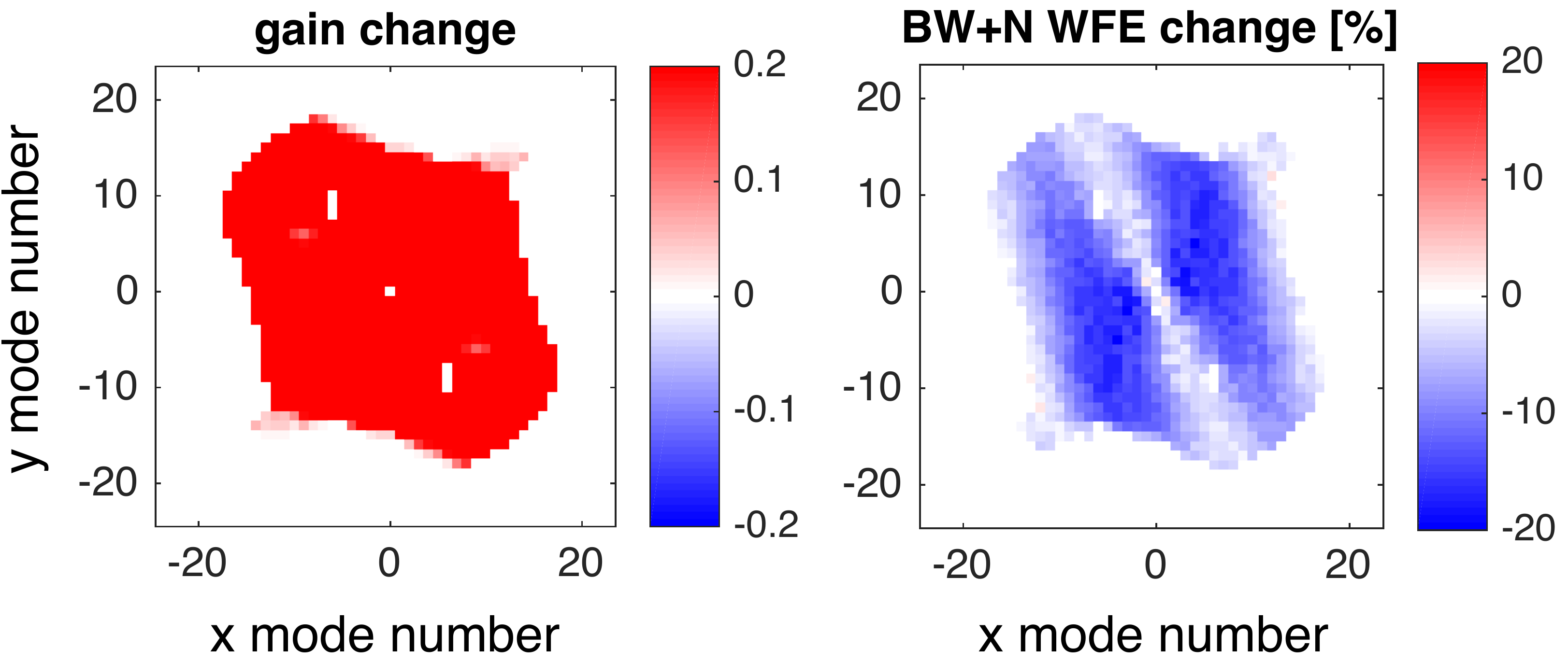}
	\end{center}
   \caption{ \label{fig:highgain_gainWFE} Maps of gain and WFE change for high gain testing. Both plots are in x/y mode space, with piston at the center. Both images are masked for clarity to show only the strongly bandwidth error-dominated region. \textit{Left:} The absolute change in gain between uniform 0.3 gains and high gains. \textit{Right:} The corresponding percent change in  bandwidth $+$ noise WFE. }
\end{figure}

As discussed in Section \ref{subsect:AO+IFS}, AO bandwidth error is the dominant factor determining raw contrast at close-to-intermediate separations, and it impacts contrasts at all separations in final PSF-subtracted data. This is particularly true for low spatial frequencies at 500~Hz loop speed, where these modes are within the controllable bandwidth of the system, but not fully corrected. In poorer seeing conditions, the system is also affected by the uncorrected centroid gain factor discussed in Section \ref{sec:intro}. Until GPI has a reliable measurement of centroid gain, we can address the symptom by increasing the control loop gain to compensate. In post-processing of AO telemetry, we can duplicate the analysis of the gain optimizer, but increase the gain cap.  The offline optimizer code is an IDL implementation of the optimization strategy described in section \ref{sect:GPIdescription}. We see in some cases (e.g. left panel of Figure \ref{fig:highgain_gainWFE}) that the optimal gains may be at least 50\% higher for some modes.

On UT April 30, 2016, we ran a high gain test on an $I=6.2$ guide star at 500~Hz. The sky transparency was constant and the ground wind speed was 4~m/s; the DIMM and MASS were not operational. Although such a bright star would normally be controlled with a 1~kHz loop, for these preliminary 500~Hz tests we chose a bright star in order to increase our signal to noise and so simplify interpretation of the results. In the high signal to noise regime, WFE (and hence IFS contrast) is dominated by bandwidth errors. We alternated GPIAO between uniform 0.3 gains and a static, non-uniform gain map with higher gains. The high gain map was calculated using the offline gain optimizer code with a gain cap of 0.5, from telemetry saved during the uniform 0.3 gain tests.

\begin{figure}[tbh]
	\begin{center}
   \includegraphics[width=6in]{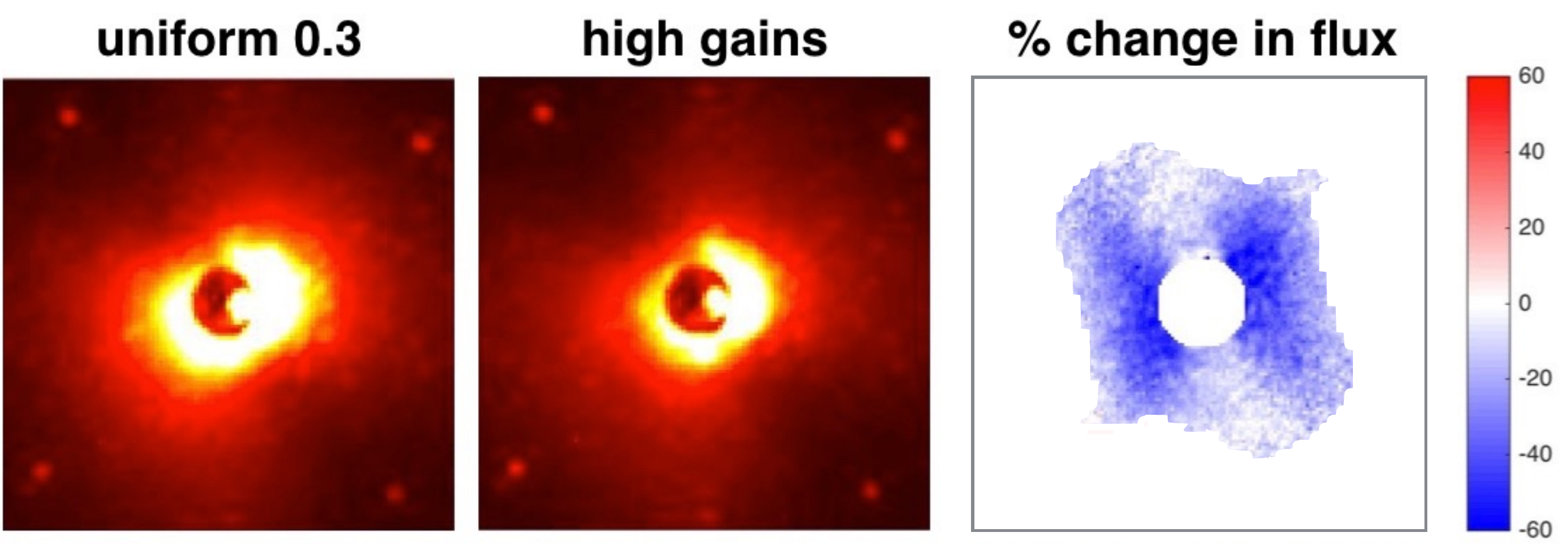}
	\end{center}
   \caption{ \label{fig:highgain_img} 1.65~$\mu$m images corresponding to the gain tests in Figure \ref{fig:highgain_gainWFE}. Image FOV is matched to the control radius. \textit{Left:} IFS image for uniform 0.3 AO gain. \textit{Middle:} IFS image for high AO gain. \textit{Right:} The percent change in flux realized by increasing the gains, masked to show the same region as in Figure \ref{fig:highgain_gainWFE}. The percent flux change is well matched to the expectation from the percent change in WFE.}
\end{figure}

Figures \ref{fig:highgain_gainWFE} and \ref{fig:highgain_img} show the gain change, WFE change, and IFS images for one pair of low/high gain tests. Note that Figure \ref{fig:highgain_gainWFE} as well as the right panel of Figure \ref{fig:highgain_img} are masked to show only the central, strongly bandwidth error-dominated, region. As anticipated, the WFE improved in the region with higher gains; the largest improvements were in low spatial frequencies along the wind direction. The IFS image contrast also improved, and the percent change in IFS flux was approximately equal to the square of the percent change in WFE, as anticipated for AO speckle-dominated regions of the PSF. In the regions of greatest improvement, the WFE decreased by 15-20\%, and the flux in the corresponding PSF regions decreased by 40\% or more. The maximum decrease in IFS flux is slightly more than we predict from the improvement in bandwidth WFE alone. This could be the result of seeing variations between the two tests, but because the DIMM and MASS were inoperable, it is not possible to independently confirm this hypothesis.

These tests show that GPI should consider increasing the gain cap on the realtime gain optimizer, particularly at 500~Hz loop speed. However, overdriving the system is a danger; the gain cap of 0.3 was chosen because it would not overdrive the system even in very good conditions (when the centroid gain is $\sim1$). An underperforming system is preferable to an unstable one, and future work will test whether the optimizer properly avoids instabilities in good conditions when the gain cap is raised.

\clearpage
\section{CONCLUSIONS}

The GPIES campaign has observed more than 300 stars in 1.5~yr, and these science and AO data provide a wealth of information about the performance of the AO system. Considerable effort has been invested in automated and semi-automated data reduction pipelines for both science data and AO telemetry, greatly facilitating performance data mining. As a result, we know that GPI AO is consistently achieving $10^{-4}$ $5\sigma$ raw contrast at 0.4'' in single 60~sec H-band images, with a stability that enables 1~hr sequences to reach a final $5\sigma$ contrast 10--20 times better.

We find several correlations between IFS images, AO telemetry, and seeing conditions. First, we find that bandwidth errors dominate the final error budget at most locations in the PSF, except for the faintest stars and/or separations beyond $\sim0.8''$. This conclusion is reinforced by the good correspondence between 2D reconstructed bandwidth WFE maps and IFS images. Interestingly,  seeing values are not well correlated with either WFE or contrast, although the atmospheric coherence time, $\tau$, is correlated. This reinforces the conclusion that GPI AO is not typically limited by the amplitude of seeing phase errors, but by the time lag relative to the speed of the turbulent layers. In a preliminary study, we used exploratory factor analysis to investigate the underlying factors responsible for raw contrast. We find four main underlying factors that track the lower atmosphere, the upper atmosphere, the AO bandwidth residual speckles, and the AO noise errors. A combination of these factors can explain 60--70\% of the variation in raw contrasts. Future work will refine and expand this analysis.

We tested the system with higher gains at 500~Hz loop speeds. Until a reliable centroid gain measurement method is implemented, GPI may consider increasing the gain cap on its gain optimizer to compensate for optical gain variations, particularly in sub-median seeing. In an on-sky test, the contrast improvement in high gain IFS images corresponded well to the expected improvement based on 2D reconstructed bandwidth WFE maps. Additional work is needed to ensure that, when operated with a higher gain cap, the gain optimizer will select stable gains in all conditions.

These analyses point to key areas for future potential upgrades. The strong dependence of performance on $\tau$ suggests GPI would benefit from wind predictive control and/or increased loop speed. Some initial studies have been done on wind measurement from GPI telemetry\cite{Srinath2016} and predictive Fourier control is being tested at ShaneAO\cite{Rudy2016}, but neither has been implemented on GPI. Several high contrast systems operate or have planned upgrades to operate at $>1$~kHz\cite{Fusco2016, Jovanovic2016, Males2016, Pinna2016}, to clear advantage with respect to lag errors.  Additionally, GPI's WFS detector noise limits the system to guide stars of $I<10$. Upgrading to an EMCCD would both increase sky coverage and allow GPI to run faster on fainter targets.

\acknowledgments
The GPI project has been supported by Gemini Observatory, which is operated by AURA, Inc., under a cooperative agreement with the NSF on behalf of the Gemini partnership: the NSF (USA), the National Research Council (Canada), CONICYT (Chile), the Australian Research Council (Australia), MCTI (Brazil) and MINCYT (Argentina).
Portions of this work were performed under the auspices of the U.S. Department of Energy by Lawrence Livermore National Laboratory under Contract DE-AC52-07NA27344.
This is document number LLNL-PROC-697077.

\bibliography{spie_library}

\begin{thebibliography}{10}

\bibitem{Macintosh2008}
Macintosh, B.~A., Graham, J.~R., Palmer, D.~W., Doyon, R., Dunn, J., Gavel,
  D.~T., Larkin, J.~E., Oppenheimer, B.~R., Saddlemyer, L., Sivaramakrishnan,
  A., Wallace, J.~K., Bauman, B., Erickson, D.~A., Marois, C., Poyneer, L.~A.,
  and Soummer, R., ``{The Gemini Planet Imager: from science to design to
  construction},'' {\em Proceedings of SPIE}~{\bf 7015},  701518 (2008).

\bibitem{Macintosh2014d}
Macintosh, B.~A., Graham, J.~R., Ingraham, P., Konopacky, Q., Marois, C.,
  Perrin, M.~D., Poyneer, L.~A., Baumann, B., Barman, T., Burrows, A.~S.,
  Cardwell, A., Chilcote, J., Rosa, R. J.~D., Dillon, D., Doyon, R., Dunn, J.,
  Erickson, D., Fitzgerald, M.~P., Gavel, D., Goodsell, S., Hartung, M., Hibon,
  P., Kalas, P.~G., Larkin, J.~E., Maire, J., Marchis, F., Marley, M.~S.,
  McBride, J., Millar-Blanchaer, M., Morzinski, K.~M., Norton, A., Oppenheimer,
  B.~R., Palmer, D.~W., Patience, J., Pueyo, L., Rantakyro, F., Sadakuni, N.,
  Saddlemyer, L., Savaransky, D., Serio, A., Soummer, R., Sivaramakrishnan, A.,
  Song, I., Thomas, S., Wallace, J.~K., Wiktorowicz, S.~J., and Wolff, S.~G.,
  ``{The Gemini Planet Imager: First Light},'' {\em Proceedings of the National
  Academy of Sciences of the United States of America}~{\bf 111}(35),  12661
  (2014).

\bibitem{Perrin2010a}
Perrin, M.~D., Graham, J.~R., Larkin, J.~E., Wiktorowicz, S.~J., Maire, J.,
  Thibault, S., Fitzgerald, M.~P., Doyon, R., Macintosh, B.~A., Gavel, D.~T.,
  Oppenheimer, B.~R., Palmer, D.~W., Saddlemyer, L., and Wallace, J.~K.,
  ``{Imaging polarimetry with the Gemini Planet Imager},'' {\em Proceedings of
  SPIE}~{\bf 7736},  77365R--77365R--9 (2010).

\bibitem{Larkin2014}
Larkin, J.~E., Chilcote, J.~K., Aliado, T., Bauman, B.~J., Brims, G., Canfield,
  J.~M., Cardwell, A., Dillon, D., Doyon, R., Dunn, J., Fitzgerald, M.~P.,
  Graham, J.~R., Goodsell, S., Hartung, M., Hibon, P., Ingraham, P., Johnson,
  C.~a., Kress, E., Konopacky, Q.~M., Macintosh, B.~a., Magnone, K.~G., Maire,
  J., McLean, I.~S., Palmer, D.~W., Perrin, M.~D., Quiroz, C., Rantakyr{\"{o}},
  F., Sadakuni, N., Saddlemyer, L., Serio, A., Thibault, S., Thomas, S.~J.,
  Vallee, P., and Weiss, J.~L., ``{The integral field spectrograph for the
  Gemini planet imager},'' {\em Proceedings of SPIE}~{\bf 9147},  91471K
  (2014).

\bibitem{Poyneer2014}
Poyneer, L.~A., {De Rosa}, R.~J., Macintosh, B.~A., Palmer, D.~W., Perrin,
  M.~D., Sadakuni, N., Savransky, D., Bauman, B., Cardwell, A., Chilcote,
  J.~K., Dillon, D., Gavel, D., Goodsell, S.~J., Hartung, M., Hibon, P.,
  Rantakyro, F.~T., Thomas, S., and Veran, J.-P., ``{On-sky performance during
  verification and commissioning of the Gemini Planet Imager's adaptive optics
  system},'' {\em Proceedings of SPIE}~{\bf 9148},  91480K (2014).

\bibitem{Poyneer2016}
Poyneer, L.~A., Palmer, D.~W., Macintosh, B.~A., Savransky, D., Sadakuni, N.,
  Thomas, S., V{\'{e}}ran, J.-P., Follette, K.~B., Greenbaum, A.~Z., {Mark
  Ammons}, S., Bailey, V.~P., Bauman, B., Cardwell, A., Dillon, D., Gavel, D.,
  Hartung, M., Hibon, P., Perrin, M.~D., Rantakyr{\"{o}}, F.~T.,
  Sivaramakrishnan, A., and Wang, J.~J., ``{Performance of the Gemini Planet
  Imager's adaptive optics system},'' {\em Applied Optics}~{\bf 55},  323 (jan
  2016).

\bibitem{Poyneer2006a}
Poyneer, L.~A., V{\'{e}}ran, J.-P., Dillon, D., Severson, S., and Macintosh,
  B.~A., ``{Wavefront control for the Gemini Planet Imager},'' {\em Proceedings
  of SPIE}~{\bf 6272},  62721N--1 (2006).

\bibitem{Poyneer2004}
Poyneer, L.~A. and Macintosh, B.~A., ``{Spatially filtered wave-front sensor
  for high-order adaptive optics},'' {\em Journal of the Optical Society of
  America A}~{\bf 21}(5),  810 (2004).

\bibitem{LeRoux2004}
{Le Roux}, B., Conan, J.-M., Kulcs{\'{a}}r, C., Raynaud, H.-F., Mugnier, L.~M.,
  and Fusco, T., ``{Optimal control law for classical and multiconjugate
  adaptive optics.},'' {\em Journal of the Optical Society of America. A}~{\bf
  21}(7),  1261--1276 (2004).

\bibitem{Poyneer2010}
Poyneer, L.~A. and V{\'{e}}ran, J.-P., ``{Kalman filtering to suppress spurious
  signals in adaptive optics control},'' {\em Journal of the Optical Society of
  America. A}~{\bf 27}(11),  A223--34 (2010).

\bibitem{Petit2014}
Petit, C., Sauvage, J.-F., Fusco, T., Sevin, A., Suarez, M., Costille, A.,
  Vigan, A., Soenke, C., Perret, D., Rochat, S., Barrufolo, A., Salasnich, B.,
  Beuzit, J.-L., Dohlen, K., Mouillet, D., Puget, P., Wildi, F., Kasper, M.,
  Conan, J.-M., Kulcs{\'{a}}r, C., and Raynaud, H.-F., ``{SPHERE eXtreme AO
  control scheme: final performance assessment and on sky validation of the
  first auto-tuned LQG based operational system},'' {\em Proceedings of
  SPIE}~{\bf 9148},  91480O (2014).

\bibitem{Sivo2014}
Sivo, G., Kulcs{\'{a}}r, C., Conan, J.-M., Raynaud, H.-F., Gendron, E., Basden,
  A., Vidal, F., Morris, T., Meimon, S., Petit, C., Gratadour, D., Martin, O.,
  Hubert, Z., Sevin, A., Perret, D., Chemla, F., Rousset, G., Dipper, N.,
  Talbot, G., Younger, E., Myers, R., Henry, D., Todd, S., Atkinson, D.,
  Dickson, C., and Longmore, A., ``{First on-sky SCAO validation of full LQG
  control with vibration mitigation on the CANARY pathfinder.},'' {\em Optics
  express}~{\bf 22}(19),  23565--23591 (2014).

\bibitem{Poyneer2005}
Poyneer, L.~A. and V{\'{e}}ran, J.-P., ``{Optimal modal Fourier-transform
  wavefront control},'' {\em Journal of the Optical Society of America A}~{\bf
  22}(8),  1515 (2005).

\bibitem{Soummer2005}
Soummer, R., ``{Apodized Pupil Lyot Coronagraphs for Arbitrary Telescope
  Apertures},'' {\em The Astrophysical Journal Letters}~{\bf 618},  L161--L164
  (2005).

\bibitem{Macintosh2016a}
Macintosh, B.~A., ``{The Gemini Planet Imager: first years on the sky},'' {\em
  Proceedings of SPIE}~{\bf 9908},  9908--19 (2016).

\bibitem{Perrin2016}
Perrin, M.~D., Ingraham, P., Follette, K.~B., Maire, J., Wang, J.~J.,
  Savransky, D., Arriaga, P., Bailey, V.~P., Bruzzone, S., Chilcote, J.~K., {De
  Rosa}, R.~J., Draper, Z., Fitzgerald, M.~P., Greenbaum, A.~Z., Hung, L.-W.,
  Konopacky, Q., Macintosh, B.~A., Marchis, F., Marois, C., Millar-Blanchaer,
  M., Nielsen, E.~L., Rameau, J., Rajan, A., Rantakyr{\"{o}}, F.~T., Ruffio,
  J.-B., Ward-Duong, K., Wolff, S.~G., and Zalesky, J., ``{Gemini Planet Imager
  Observational Calibrations XI : Pipeline Improvements and Enhanced
  Calibrations after Two Years On Sky},'' {\em Proceedings of SPIE}~{\bf 9908},
   9908--121 (2016).

\bibitem{Sarazin1990}
Sarazin, M. and Roddier, F., ``{The ESO differential image motion monitor},''
  {\em Astronomy {\&} Astrophysics}~{\bf 227},  294--300 (1990).

\bibitem{Kornilov2003}
Kornilov, V., Tokovinin, A., Vozyakova, O., Zaitsev, A., Shatsky, N., Potanin,
  S., and Sarazin, M., ``{MASS: a monitor of the vertical turbulence
  distribution},'' {\em Proceedings of SPIE}~{\bf 4839},  837--845 (2003).

\bibitem{Marois2006}
Marois, C., Lafreni{\`{e}}re, D., Doyon, R., Macintosh, B.~A., and Nadeau, D.,
  ``{Angular differential imaging: a powerful high-contrast imaging
  technique},'' {\em The Astrophysical Journal}~{\bf 641},  556--564 (2006).

\bibitem{Soummer2012}
Soummer, R., Pueyo, L., and Larkin, J.~E., ``{Detection and Characterization of
  Exoplanets and Disks Using Projections on Karhunen-Lo{\`{e}}ve
  Eigenimages},'' {\em The Astrophysical Journal Letters}~{\bf 755},  L28 (aug
  2012).

\bibitem{Wang2015b}
Wang, J.~J., Ruffio, J.-B., {De Rosa}, R.~J., Aguilar, J., Wolff, S.~G., and
  Pueyo, L.~A., ``{PyKLIP},'' {\em Astrophysics Source Code Library}~{\bf
  1506.001} (2015).

\bibitem{Poyneer2006}
Poyneer, L.~A. and Macintosh, B.~A., ``{Optimal Fourier control performance and
  speckle behavior in high-contrast imaging with adaptive optics},'' {\em
  Optics Express}~{\bf 14},  7499--7514 (2006).

\bibitem{IBM2015a}
{IBM Corp.}, ``{SPSS Statistics for Macintosh v23.0},'' (2015).

\bibitem{Srinath2016}
Srinath, S., Poyneer, L.~A., Gavel, D.~T., Rudy, A.~R., and Ammons, S.~M.,
  ``{Creating the right atmosphere: analyzing extensive telemetry data to
  create more realistic, computationally efficient atmosphere models for
  end-to-end simulations},'' {\em Proceedings of SPIE}~{\bf 9911},  9911--50
  (2016).

\bibitem{Rudy2016}
Rudy, A.~R., Srinath, S., Poyneer, L.~A., Gavel, D.~T., and Ammons, S.~M.,
  ``{Demonstration of predictive Fourier control for frozen flow wind
  turbulence on the ShaneAO 3-meter AO system},'' {\em Proceedings of
  SPIE}~{\bf 9909},  9909--23 (2016).

\bibitem{Fusco2016}
Fusco, T., Beuzit, J.-L., Sauvage, J.-F., Mouillet, D., Kasper, M.~E., Petit,
  C., Dohlen, K., Costille, A., Girard, J. H.~V., {Su{\'{a}}rez Valle}, M.,
  Soenke, C., Hubin, N., Milli, J., Vigan, A., Baruffolo, A., Salasnich, B.,
  Zins, G., Perret, D., Baudoz, P., and Wildi, F., ``{SAXO, the SPHERE extreme
  AO system: on-sky final performance and future improvements},'' {\em
  Proceedings of SPIE}~{\bf 9909},  9909--32 (2016).

\bibitem{Jovanovic2016}
Jovanovic, N., Guyon, O., Lozi, J., Currie, T., Hagelberg, A., Norris, B.~R.,
  Singh, G., Pathak, P., Doughty, D., Goebel, S.~B., Males, J.~R., Kuhn, J.~G.,
  Serabyn, E., Tuthill, P.~G., Schworer, G., Martinache, F., Kudo, T.,
  Kawahara, H., Kotani, T., Ireland, M.~J., Feger, T., Rains, A.~D., Bento, J.,
  Schwab, C., Coutts, D.~W., Cvetojevic, N., Gross, S., Arriola, A., Lagadec,
  T., Kasdin, J., Groff, T., Mazin, B., Minowa, Y., Takato, N., and Tamura, M.,
  ``{The SCExAO high contrast imager: transitioning from commissioning to
  science},'' {\em Proceedings of SPIE}~{\bf 9909},  9909--34 (2016).

\bibitem{Males2016}
Males, J.~R., Close, L.~M., Morzinski, K.~M., Guyon, O., and Hinz, P.~M.,
  ``{The path to visible extreme adaptive optics with MagAO},'' {\em
  Proceedings of SPIE}~{\bf 9909},  9909--184 (2016).

\bibitem{Pinna2016}
Pinna, E., Esposito, S., Hinz, P.~M., Agapito, G., Bonaglia, M., Carbonaro, L.,
  Puglisi, A.~T., Xompero, M., Riccardi, A., Montoya, O., and Durney, O.,
  ``{SOUL: the Single conjugated adaptive Optics Upgrade for LBT},'' {\em
  Proceedings of SPIE}~{\bf 9909},  9909--126 (2016).

\end{thebibliography}
\bibliographystyle{spiebib} 

\end{document}